\documentclass[%
 aip,
 sd,%
 amsmath,amssymb,
 reprint,%
]{revtex4-1}

\usepackage{graphicx}
\usepackage{dcolumn}
\usepackage{bm}
\usepackage{tabularx}

\graphicspath{ {./figs/} }

\begin{document}

\preprint{AIP/123-QED}

\title{Terahertz intersubband absorption in non-polar m-plane AlGaN/GaN quantum wells}


\author{C. Edmunds}
\affiliation{Physics Department, Purdue University, West Lafayette, Indiana 47907, USA}

\author{J. Shao}
\affiliation{Physics Department, Purdue University, West Lafayette, Indiana 47907, USA}
\affiliation{Birck Nanotechnology Center, Purdue University, West Lafayette, Indiana 47907, USA}

\author{M. Shirazi-HD}
\affiliation{Birck Nanotechnology Center, Purdue University, West Lafayette, Indiana 47907, USA}
\affiliation{School of Electrical and Computer Engineering, Purdue University, West Lafayette, Indiana 49707, USA}

\author{M. J. Manfra}
\affiliation{Physics Department, Purdue University, West Lafayette, Indiana 47907, USA}
\affiliation{Birck Nanotechnology Center, Purdue University, West Lafayette, Indiana 47907, USA}
\affiliation{School of Electrical and Computer Engineering, Purdue University, West Lafayette, Indiana 49707, USA}
\affiliation{School of Materials Engineering, Purdue University, West Lafayette, Indiana 49707, USA}

\author{O. Malis}
\email[]{omalis@purdue.edu}
\affiliation{Physics Department, Purdue University, West Lafayette, Indiana 47907, USA}

\date{\today}

\begin{abstract}
We demonstrate THz intersubband absorption (15.6-26.1 meV) in m-plane AlGaN/GaN quantum wells. We find a trend of decreasing peak energy with increasing quantum well width, in agreement with theoretical expectations. However, a blue-shift of the transition energy of up to 14 meV was observed relative to the calculated values. This blue-shift is shown to decrease with decreasing charge density and is therefore attributed to many-body effects. Furthermore, a $\sim$40\% reduction in the linewidth (from roughly 8 to 5 meV) was obtained by reducing the total sheet density and inserting undoped AlGaN layers that separate the wavefunctions from the ionized impurities in the barriers. 
\end{abstract}



\maketitle 

\begin{figure}
	\includegraphics[width=0.5\textwidth]{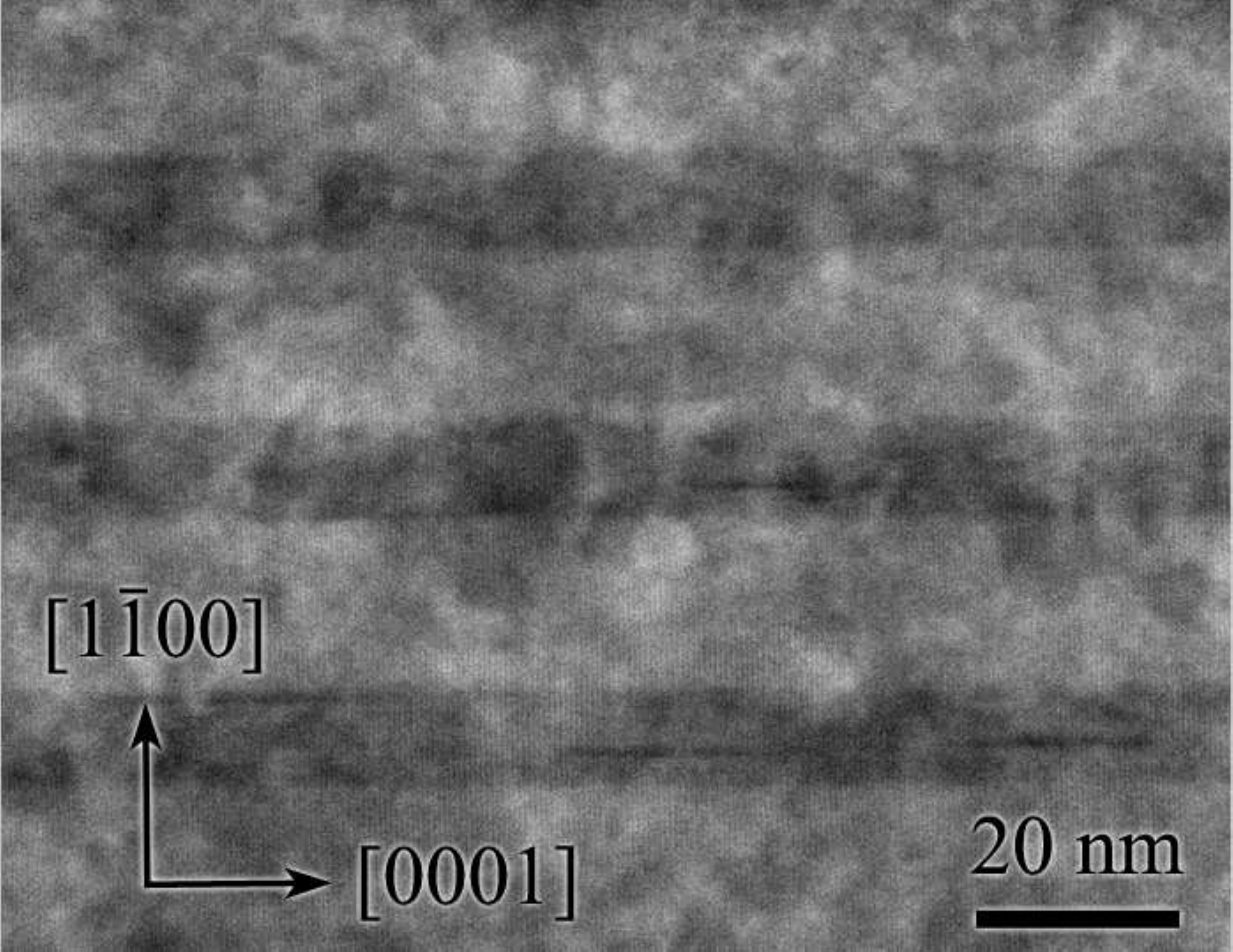}
	\caption{\label{fig:stemC} HAADF-STEM image of sample C.}
\end{figure}

\begin{figure}
	\includegraphics[width=0.5\textwidth]{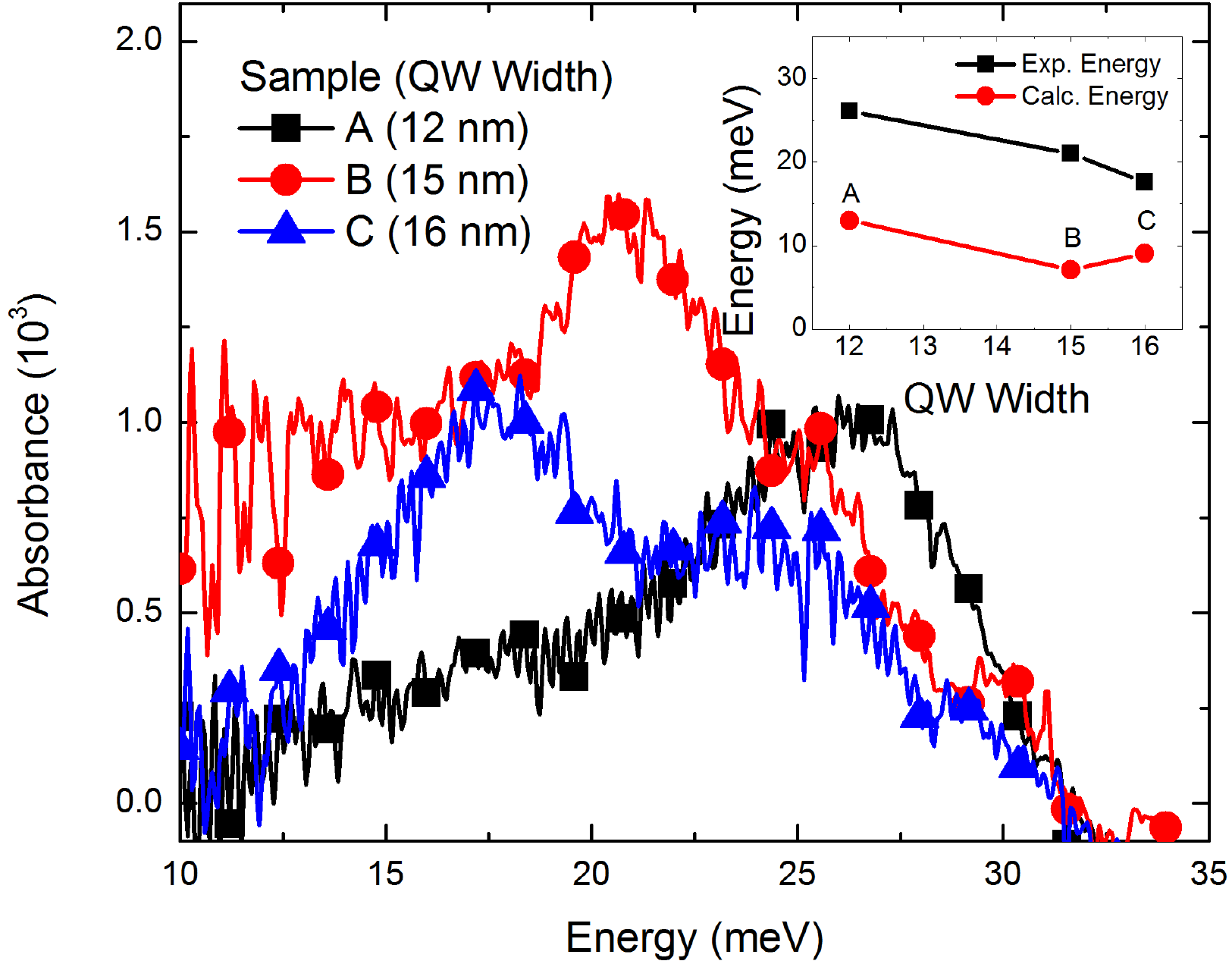}
	\caption{\label{fig:absAC} (Color online) Absorbance spectra for samples A through C measured at 9 K. Symbols are displayed every 20 data points. Inset displays the experimental and calculated 1$\rightarrow$2 transition energies as a function of QW width. Text indicates the sample associated with a given data point.}
\end{figure}

The terahertz (THz) spectral region has attracted attention due to potential applications in medical diagnostics, security screening and quality control. GaAs/AlGaAs quantum cascade lasers (QCLs) have already demonstrated potential as THz sources in the 1.2-5 THz range.\cite{Koehler:2002,Williams:2005,Williams:2006,Walther:2007,Luo:2007} However, the operating range of GaAs QCLs is limited by the longitudinal optical (LO) phonon emission at 36 meV (8.7 THz). Fortunately, GaN-based QCLs have the potential to operate in this range due to the larger LO-phonon energy (90 meV).
\\{\indent}To date, most studies of intersubband transitions in the III-nitrides have utilized the polar c-plane orientation.\cite{Tchernycheva:2006,Kandaswamy:2008,Malis:2009,Machhadani:2010,Edmunds:2013} Spontaneous emission from c-plane AlGaN/GaN QCLs in the THz region has been reported, although full laser operation has remained elusive.\cite{Terashima:2011} The built-in polarization fields in c-plane heterostructures place a lower limit on the transition energy, and the inherent asymmetry in the conduction band profile reduces the dipole moment at the larger well widths required for operation in the THz region. These limitations have been partially mitigated by the implementation of more complex step-well designs.\cite{Machhadani:2010,Sudradjat:2012} However, the transition energies of these step-wells are highly sensitive to structural parameters.\cite{Wu:2013,Beeler:2013a} Moreover, the additional layers significantly increase the complexity of design and growth of practical devices. The challenges of the built-in polarization fields can be circumvented by utilizing non-polar nitride heterostructures. Non-polar nitride structures may be achieved using either the cubic phase or the m-plane orientation of the wurzite phase. Near-infrared intersubband absorption in the cubic AlGaN/GaN has been observed,\cite{Machhadani:2011} and m-plane oriented quantum well infrared photodetectors operating in the mid-infrared have recently been demonstrated.\cite{Pesach:2013} However, THz intersubband transitions in non-polar nitrides have not yet been reported. This paper investigates the intersubband absorption properties of non-polar m-plane AlGaN/GaN quantum wells in the THz region.
\\{\indent} We have already demonstrated molecular beam epitaxy (MBE) growth of high-quality m-plane AlGaN/GaN superlattices. \cite{Shao:2013a,Shao:2013b} In this study, the samples consist of 26 quantum wells (QWs) grown on free-standing m-plane GaN substrates from Kyma Technologies at 720$^{\circ}$C. The well width was varied from 12 to 16 nm to verify the expected dependence of the intersubband peak energy on well width. For samples A through C (see Table \ref{ta:THzsamples}), the barriers consists of 10 nm Al$_{x}$Ga$_{1-x}$N layers with an Al composition of either 4 or $16\%$ doped with Si at $1 \times 10^{18}$ cm$^{-3}$. For samples D and E, the barrier thickness was increased to 13.5 nm. For the latter samples, the barriers consist of two 4.5 nm undoped spacer regions, and a 4.5 nm central region with Si concentration of $1 \times 10^{18}$ cm$^{-3}$. The layer thickness and alloy composition were determined using high-resolution x-ray diffraction (HR-XRD), assuming equal growth rates for the well and barrier materials.\cite{Shao:2013b} To verify the layer thickness and assess interface roughness, high-angle annular dark-field scanning transmission electron microscopy (HAADF-STEM) imaging was carried out for sample C (Figure \ref{fig:stemC}). The layer thickness of the well and barrier material were found to be 16.6 and 10.0 nm, respectively, in agreement with the values obtained from HR-XRD to within a few monolayers (see Table \ref{ta:THzsamples}).
\begin{table}[b!]
  \caption{\label{ta:THzsamples} QW and barrier width, Al composition, experimental and calculated transition energy, and full-width-at-half-maximum for the samples in this study.}
  \begin{tabularx}{0.5\textwidth}{@{}l *6{>{\centering\arraybackslash}X}@{}}
  \hline
  Sample&QW Width (nm)&Barrier Width (nm)&Al comp. (\%)&Exp. Energy (meV)&Calc. Energy (meV)&FWHM (meV)\\
  \hline
  \hline
  A&12.0&10.0&4.0&26.1&13.0&9.0\\
  B&15.0&10.0&3.8&21.0&7.1&7.0\\
  C&16.0&10.0&16&17.6&9.5&8.0\\
  D&14.5&13.5&16&20.6&16.8&4.5\\
  E&16.1&13.5&16&15.6&13.0&5.3\\
  \end{tabularx}
\end{table}
\\{\indent}The optical properties of the samples were characterized using Fourier-transform infrared spectroscopy, utilizing a liquid helium cooled Si bolometer. A polyethylene grid polarizer was used to isolate the intersubband transitions. This setup provides a measurement range from roughly 10 to 35 meV. This range is limited at higher energies by onset of the GaN phonon band \cite{Yang:2005} and at lower energies by the signal-to-noise ratio of our experimental setup. Samples were polished into 45$^\circ$ multipass waveguides and measured in a liquid helium cooled continuous flow cryostat at 9 K. Due to the electro-magnetic boundary conditions, a node in the electric-field is expected at the air-semiconductor interface.\cite{Helm:1999} The impact of this boundary condition on the absorption properties in the near-infrared is relatively small due to the shorter wavelengths involved (1-3 $\mu$m). However, at the longer wavelengths involved in this study, the overlap of the electric-field with the QWs is expected to be negligible. The boundary conditions can be reversed by placing a metal at the QW surface. Therefore, to enhance the overlap of the electric-field with the QWs, a gold film (200 nm) was deposited on the surface via thermal evaporation. Under these conditions, a maximum in the electric-field overlap is expected.\cite{Helm:1999}
\\{\indent}The band structure of the AlGaN/GaN QWs was calculated self-consistently using the effective mass model and the \textbf{nextnano++} software.\cite{Birner:2007} The effects of non-parabolicity were ignored due to the small transition energies. The material parameters used in the calculation are summarized in Reference 22\nocite{Wu:2009}. Assuming a linear interpolation of AlN and GaN bandgap parameters, conduction band offsets of 82 and 326 meV were calculated for Al$_{0.04}$Ga$_{0.96}$N/GaN and Al$_{0.16}$Ga$_{0.84}$N/GaN, respectively. The dopants were considered to be fully ionized. Many-body effects have been observed previously in the nitrides.\cite{Tchernycheva:2006,Kandaswamy:2008,Kandaswamy:2010,Edmunds:2012b} However, due to the small transition energies observed in this study, many-body effects cannot be accurately estimated using the standard perturbative approach and were ignored in the calculation.
\\{\indent}Fig. \ref{fig:absAC} displays the absorbance curves for samples A through C.
The absorbance spectra were extracted from the ratio of the p- and s-polarized transmission measurements. To account for the experimental geometry, the absorbance spectra were then normalized by the number of passes through the active region, the number of QWs, polarization coupling angle (45$^\circ$) and electric-field overlap (2). For samples A and B, a polarization sensitive peak was observed at 26.1 and 21.0 meV, respectively. These peaks are attributed to the 1$\rightarrow$2 transition in the QWs. For sample C, two peaks were observed: a dominant peak at 17.6 meV, which is attributed to the 1$\rightarrow$2 transition and a secondary feature at 25.0 meV, which we associate with a higher order transition made possible by the larger well width and barrier height (composition). The 1$\rightarrow$2 transition energy in each sample follows the trend of decreasing energy with increasing QW width. However, the experimental values are systematically higher than expected (see Fig. \ref{fig:absAC}, inset).
\\{\indent}The higher experimental transition energies are attributed to many-body effects. However, these effects cannot be treated with the perturbative approach\cite{Tchernycheva:2006} since the transition energies (18-26 meV) are comparable to the many-body shift ($\sim$10 meV). The total sheet density for samples A-C is $1 \times 10^{12}$ cm$^{-2}$. At these sheet densities, the standard perturbative approach to many-body corrections yields a shift of roughly 50 meV, clearly too large to reproduce the experimental results. A more rigorous solution of the Hartree-Fock equation may be required to accurately determine the impact of many-body effects.\cite{Kandaswamy:2010}
\begin{figure}[t]
	\includegraphics[width=0.5\textwidth]{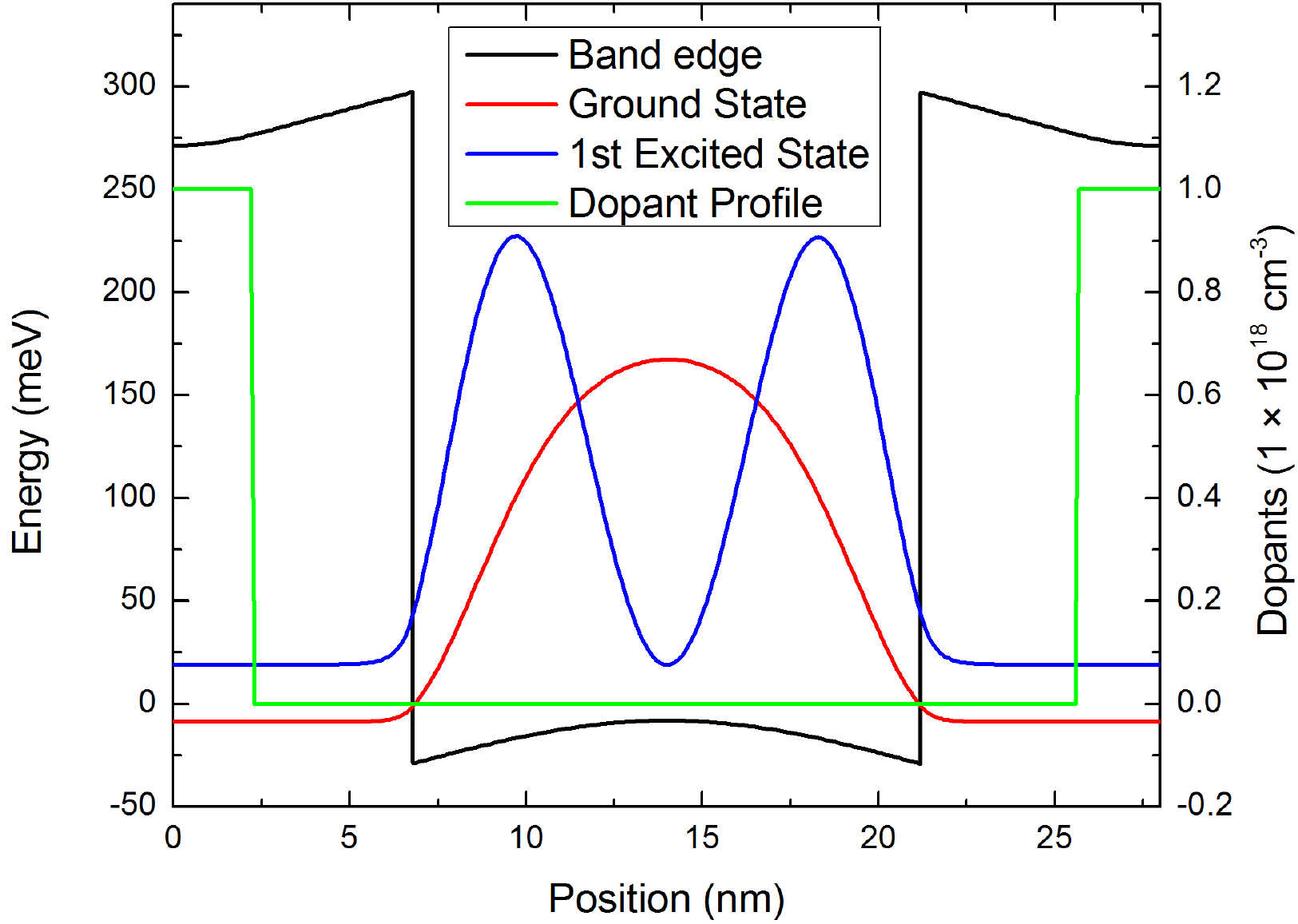}
	\caption{\label{fig:dopmod} (Color online) Band edge, wavefunctions and doping profile for the sample D.}
\end{figure}
\\{\indent}A full-width-at-half-maximum (FWHM) of 7-9 meV was observed for samples A-C (see Table \ref{ta:THzsamples}). This linewidth is sufficient to cover a large portion of the measurement range. Due to the low transition energies and measurement temperature (9 K), phonon scattering is suppressed. Furthermore, we do not expect interface roughness to contribute significantly to the linewidth since sharp interfaces were observed in HAADF-STEM images (Figure \ref{fig:stemC}). Based on calculations of the intersubband lifetime in the near-infrared,\cite{Edmunds:2013} we expect ionized impurity scattering to be the dominant contribution to the linewidth.
\begin{figure}
	\includegraphics[width=0.5\textwidth]{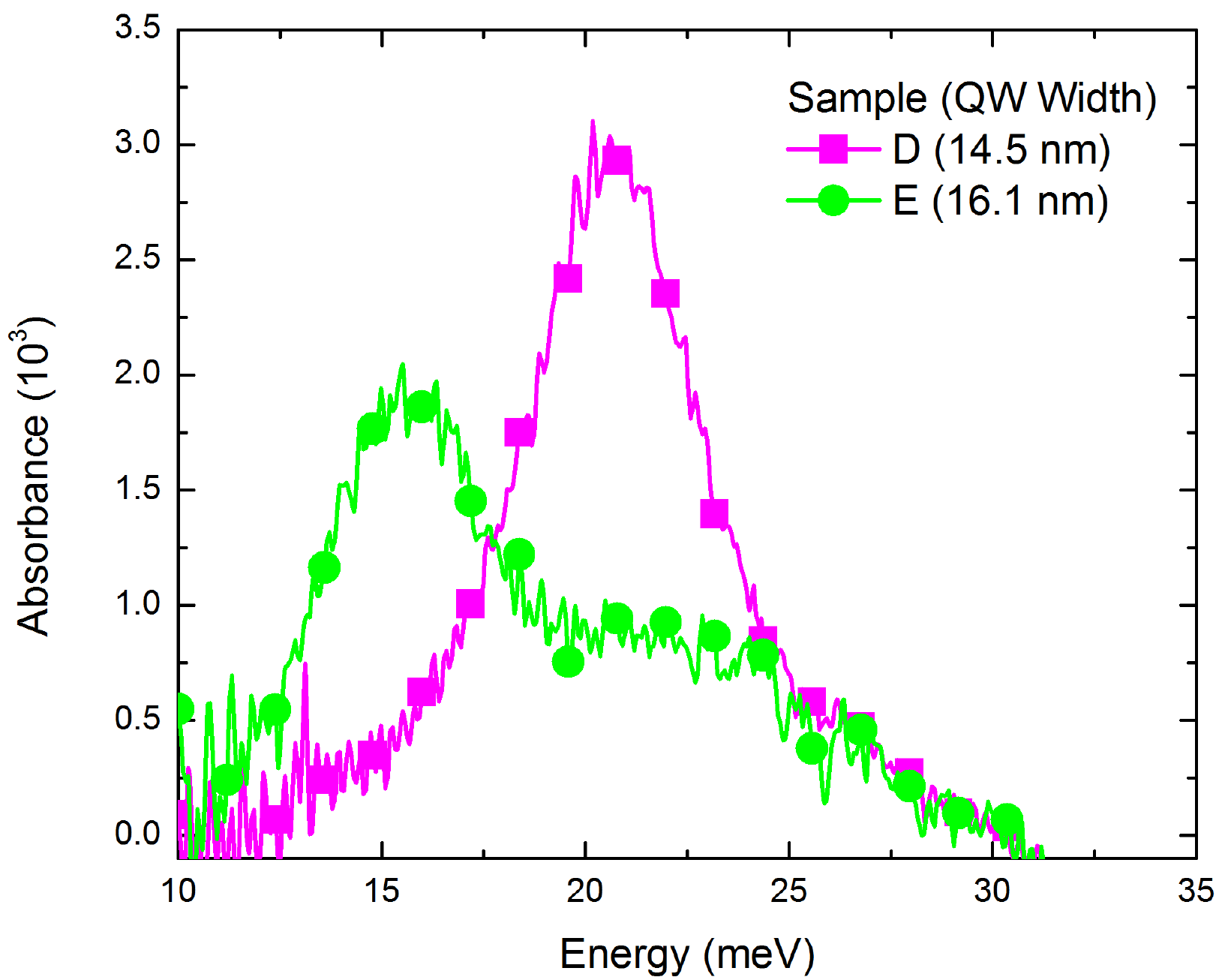}
	\caption{\label{fig:absDE} (Color online) Absorbance spectra for samples D and E measured at 9 K. Symbols are displayed every 20 data points. }
\end{figure}
\\{\indent}To investigate the impact of ionized impurity scattering on linewidth, samples D and E were grown with partially doped barriers. These 13.5 nm barriers consist of two 4.5 nm undoped spacer regions, and a 4.5 nm central region with Si concentration of $1 \times 10^{18}$ cm$^{-3}$ (see Fig. \ref{fig:dopmod}). This design is expected to reduce the overlap of the wavefunctions with ionized impurities. Furthermore, the reduced width of the doped region in samples D and E compared to samples A-C results in a reduction of the total sheet density from $1 \times 10^{12}$ to $4.5 \times 10^{11}$ cm$^{-2}$. This reduction in total charge density is expected to reduce carrier-carrier scattering, and the impact of many-body effects.
\\{\indent}The absorbance spectra for samples D and E are shown in Figure \ref{fig:absDE}. A significant decrease of the FWHM was observed from roughly 8 meV (samples A-C) to 5 meV (samples D and E). Therefore, we attribute the large linewidth of samples A-C primarily to ionized impurity and carrier-carrier scattering. Furthermore, the 1$\rightarrow$2 transition energy of 20.6 meV (15.6 meV) for sample D (E) is in reasonable agreement with the calculated value of 16.8 meV (13.0 meV). A higher energy peak is also observed for sample E at 22.5 meV, which we attribute to a higher order transition. The absence of any higher order transition in the spectrum of sample D may be due to the limited range of the experiment. The blue shift of the 1$\rightarrow$2 transition energy for samples D and E (up to 3.8 meV) is significantly reduced compared to samples A-C (up to 14 meV). This reduction in the blue shift is consistent with the reduction of the total charge density. 
\\{\indent}The integrated absorbance characterizes the transition strength, and is determined by integrating over the absorbance spectrum. For the samples measured in this study, the experimental values of the integrated absorbance varied from roughly 10 to 20 $\mu$eV. We note that these values are significantly lower than expected (60-160 $\mu$eV). This difference is attributed to deviation of the electric-field distribution from the ideal boundary conditions assumed during normalization. A more accurate calculation must consider the variation in refractive index of GaN due to the presence of the QWs and nearby phonon absorption bands.\cite{Zaluzny:1999}
\\{\indent} In summary, we have demonstrated THz intersubband absorption (15.6-26.1 meV) in m-plane AlGaN/GaN QWs. For the 1$\rightarrow$2 transition, we observe a trend of decreasing energy with increasing QW width, in agreement with theoretical expectations. However, a many-body blue-shift of up to 14 meV was observed relative to the calculated values. This blue-shift is shown to decrease to roughly 4 meV when the total sheet density is reduced from $1 \times 10^{12}$ cm$^{-2}$ (samples A-C) to $4.5 \times 10^{11}$ cm$^{-2}$ (samples D and E). Furthermore, a $\sim$40\% reduction in the linewidth (from roughly 8 to 5 meV) was obtained by reducing the total sheet density and inserting an undoped spacer region that separates the wavefunctions from the ionized impurities in the barriers. Our results are promising for the potential use of m-plane nitrides for THz lasers and detectors.
\\{\indent}This work was supported by the NSF awards ECCS-1001431, ECCS-1253720, and DMR-1206919.


%
%

%


\bibliography{mplaneTHzabs}

\end{document}